\documentclass{amsart}
\usepackage[dvips]{graphicx}
\usepackage{graphicx}
\usepackage{amscd}
\usepackage{amsmath}
\usepackage{amsxtra}
\usepackage{amsfonts}
\usepackage{amssymb}

\newtheorem{theorem}{Theorem}[section]

\newtheorem{lemma}[theorem]{Lemma}
\newtheorem{proposition}[theorem]{Proposition}
\theoremstyle{definition}
\newtheorem{definition}[theorem]{Definition}
\newtheorem{remark}[theorem]{Remark}

\newtheorem{example}[theorem]{Example}
\theoremstyle{remark}

\renewcommand{\theclaim}{\textup{\theclaim}}

\newtheorem*{acknowledgements}{Acknowledgements}

\numberwithin{equation}{section}

\def\openone

{\mathchoice

{\hbox{\upshape \small1\kern-3.3pt\normalsize1}}

{\hbox{\upshape \small1\kern-3.3pt\normalsize1}}

{\hbox{\upshape \tiny1\kern-2.3pt\SMALL1}}

{\hbox{\upshape \Tiny1\kern-2pt\tiny1}}}

\makeatletter

\newbox\ipbox

\newcommand{\diracb}[1]{\left\langle #1\mathrel{\mathchoice

{\setbox\ipbox=\hbox{$\displaystyle \left\langle\mathstrut
#1\right.$}

\vrule height\ht\ipbox width0.25pt depth\dp\ipbox}

{\setbox\ipbox=\hbox{$\textstyle \left\langle\mathstrut
#1\right.$}

\vrule height\ht\ipbox width0.25pt depth\dp\ipbox}

{\setbox\ipbox=\hbox{$\scriptstyle \left\langle\mathstrut
#1\right.$}

\vrule height\ht\ipbox width0.25pt depth\dp\ipbox}

{\setbox\ipbox=\hbox{$\scriptscriptstyle \left\langle\mathstrut
#1\right.$}

\vrule height\ht\ipbox width0.25pt depth\dp\ipbox}

}\right. }

\newcommand{\dirack}[1]{\left. \mathrel{\mathchoice

{\setbox\ipbox=\hbox{$\displaystyle \left.\mathstrut
#1\right\rangle$}

\vrule height\ht\ipbox width0.25pt depth\dp\ipbox}

{\setbox\ipbox=\hbox{$\textstyle \left.\mathstrut
#1\right\rangle$}

\vrule height\ht\ipbox width0.25pt depth\dp\ipbox}

{\setbox\ipbox=\hbox{$\scriptstyle \left.\mathstrut
#1\right\rangle$}

\vrule height\ht\ipbox width0.25pt depth\dp\ipbox}

{\setbox\ipbox=\hbox{$\scriptscriptstyle \left.\mathstrut
#1\right\rangle$}

\vrule height\ht\ipbox width0.25pt depth\dp\ipbox}

} #1\right\rangle}

\usepackage{graphicx}

\def\blfootnote{\xdef\@thefnmark{}\@footnotetext}

\begin{document}
\title[Entropy Encoding, Hilbert Space, and Karhunen-Lo\`{e}ve Transforms]{Entropy Encoding, Hilbert Space and Karhunen-Lo\`{e}ve Transforms}
\author{Palle E.T. Jorgensen}
\address[Palle E.T. Jorgensen]{Department of Mathematics\\
The University of Iowa\\
14 MacLean Hall\\
Iowa City, IA 52242}
\email{jorgen@math.uiowa.edu}

\author{Myung-Sin Song}
\address[Myung-Sin Song]{Department of Mathematics and Statistics\\
Southern Illinois University Edwardsville\\
Box 1653, Science Building\\
Edwardsville, IL 62026}
\email{msong@siue.edu}\ 

\thanks{Work supported in part by the U.S. National Science
Foundation}
\begin{abstract}
By introducing Hilbert space and operators, we show how probabilities, approximations and entropy encoding from signal and image processing allow precise formulas and quantitative estimates.  Our main results yield orthogonal bases which optimize distinct measures of data encoding.
\end{abstract}

\maketitle \tableofcontents
\section{Introduction}
\label{sec:1}

Historically, the Karhunen-Lo\`{e}ve arose as a tool from the interface of probability theory and information theory; see details with references inside the paper. It has served as a powerful tool in a variety of applications; starting with the problem of separating variables in stochastic processes, say $X_{t}$; processes that arise from statistical noise, for example from fractional Brownian motion. Since the initial inception in mathematical statistics, the operator algebraic contents of the arguments have crystallized as follows: starting from the process $X_{t}$,  for simplicity assume zero mean, i.e., $E(X_{t}) = 0$; create a correlation matrix  $C(s,t) = E(X_{s} X_{t})$. (Strictly speaking, it is not a matrix, but rather an integral kernel. Nonetheless, the matrix terminology has stuck.) The next key analytic step in the Karhunen-Lo\`{e}ve method is to then apply the Spectral Theorem from operator theory to a corresponding selfadjoint operator, or to some operator naturally associated with $C$: Hence the name, the Karhunen-Lo\`{e}ve Decomposition (KLC). In favorable cases (discrete spectrum), an orthogonal family of functions $(f_{n}(t))$ in the time variable arise, and a corresponding family of eigenvalues. We take them to be normalized in a suitably chosen square-norm. By integrating the basis functions $f_{n}(t)$ against $X_{t}$, we get a sequence of random variables $Y_{n}$. It was the insight of Karhunen-Lo\`{e}ve \cite{Loe52} to give general conditions for when this sequence of random variables is independent, and to show that if the initial random process $X_{t}$ is Gaussian, then so are the random variables $Y_{n}$. (See also Example 3.1 below.)

In the 1940s, Kari Karhunen (\cite{Kar46}, \cite{Kar52}) pioneered the use of spectral theoretic methods in the analysis of time series, and more generally in stochastic processes. It was followed up by papers and books by Michel Lo{\`e}ve in the 1950s \cite{Loe52}, and in 1965 by R.B. Ash \cite{Ash90}. (Note that this theory precedes the surge in the interest in wavelet bases!) 

As we outline below, all the settings place rather stronger assumptions. We argue how more modern applications dictate more general theorems, which we prove in our paper. A modern tool from operator theory and signal processing which we will use is the notion of \textit{frames} in Hilbert space. More precisely, frames are redundant ``bases" in Hilbert space. They are called framed, but intuitively should be thought of as generalized bases. The reason for this, as we show, is that they offer an explicit choice of a (non-orthogonal) expansion of vectors in the Hilbert space under consideration.

In our paper, we rely on the classical literature (see e.g., \cite{Ash90}), and we accomplish three things: (i) We extend the original Karhunen-Lo\`{e}ve idea to case of continuous spectrum; (ii) we give frame theoretic uses of the Karhunen-Lo\`{e}ve idea which arise in various wavelet contexts and which go beyond the initial uses of  Karhunen-Lo\`{e}ve; and finally (iii) to give applications.

These applications in our case come from image analysis; specifically from the problem of statistical recognition and detection; e.g., to nonlinear variance, for example due to illumination effects. Then the Karhunen-Lo\`{e}ve Decomposition (KLD), also known as Principal Component Analysis (PCA) applies to the intensity images. This is traditional in statistical signal detection and in estimation theory. Adaptations to compression and recognition are of a more recent vintage. In brief outline, each intensity image is converted into vector form. (This is the simplest case of a purely intensity-based coding of the image, and it is not necessarily ideal for the application of KL-decompositions.)

The ensemble of vectors used in a particular conversion of images is assumed to have a multi-variate Gaussian distribution since human faces form a dense cluster in image space. The PCA method generates small set of basis vectors forming subspaces whose linear combination offer better (or perhaps ideal) approximation to the original vectors in the ensemble. In facial recognition, the new bases are said to span intra-face and inter-face variations, permitting Euclidean distance measurements to exclusively pick up changes in for example identity and expression. 

Our presentation will start with various operator theoretic tools, including frame representations in Hilbert space. We have included more details and more explanations than is customary in more narrowly focused papers, as we wish to cover the union of four overlapping fields of specialization, operator theory, information theory, wavelets, and physics applications.

While entropy encoding is popular in engineering, \cite{SkChEb01}, 
\cite{Use01}, \cite{DoVeDeDa98} the choices made in signal processing are 
often more by trial and error than by theory.  Reviewing the literature, we 
found that the mathematical foundation of the current use of entropy in 
encoding deserves closer attention.

In this paper we take advantage of the fact that Hilbert space and operator 
theory form the common language of both quantum mechanics and of signal/image 
processing. Recall first that in quantum mechanics, (pure) states as 
mathematical entities ``are" one-dimensional subspaces in complex Hilbert space
 $\mathcal{H}$, so we may represent them by vectors of norm one. Observables 
``are" selfadjoint operators in $\mathcal{H}$, and the measurement problem entails von 
Neumann's spectral theorem applied to the operators.

In signal processing, time-series, or matrices of pixel 
numbers may similarly be realized by vectors in Hilbert space $\mathcal{H}$. The 
probability distribution of quantum mechanical observables (state space 
$\mathcal{H}$) may be represented by choices of orthonormal bases (ONBs) in 
$\mathcal{H}$ in the usual way (see e.g., \cite{Jor06}). In signal/image 
processing, because of aliasing, it is practical to generalize the notion of 
ONB, and this takes the form of what is called ``a system of frame vectors''; 
see \cite{Chr03}. 

But even von Neumann's measurement problem, viewing experimental data as part 
of a bigger environment (see e.g., \cite{DS97}, \cite{Wa04}, \cite{El03}) leads
 to basis notions more general than ONBs. They are commonly known as Positive 
Operator Valued Measures (POVMs), and in the present paper we examine the 
common ground between the two seemingly different uses of operator theory in 
the separate applications. To make the paper presentable to two audiences, we 
have included a few more details than is customary in pure math papers.

We show that parallel problems in quantum mechanics and in signal processing 
entail the choice of ``good" orthonormal bases (ONBs). One particular such ONB 
goes under the name ``the Karhunen-Lo\`{e}ve basis." We will show that it is 
optimal in three ways, and we will outline a number of applications.

The problem addressed in this paper is motivated by consideration of the optimal choices of bases for certain analogue-to-digital (A-to-D) problems we encountered in the use of wavelet bases in image-processing 
(see \cite{GoWoEd04}, \cite{SkChEb01}, \cite{Use01}, \cite{Wal99}); but certain of our considerations have an operator theoretic flavor which we wish to isolate, as it seems to be of independent interest.

There are several reasons why we take this approach. Firstly our Hilbert space 
results seem to be of general 
interest outside the particular applied context where we encountered them. And 
secondly, we feel that our more abstract results might inspire workers in 
operator theory and approximation theory. 

\subsection{Digital Image Compression}
\label{sec:DIC}
\begin{figure}[htb]
\label{F:outline}
  \begin{center}
    \includegraphics[width=5.25in]{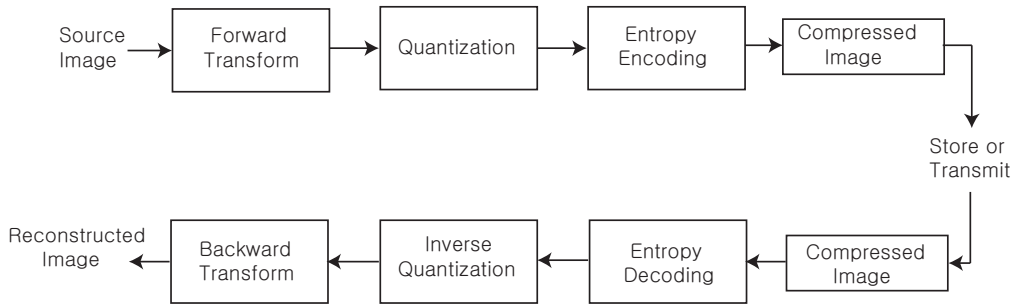}
    \caption{Outline of the wavelet image compression process. \cite{SkChEb01}}
  \end{center}
\end{figure}

In digital image compression, after the quantization (see Figure 
\ref{F:outline}) entropy encoding 
is performed on a particular image for more efficient-less storage 
memory-storage.  When an image is to be stored we need either 8 bits or 16 
bits to store a pixel.  With efficient entropy encoding, we can use a smaller 
number of bits to represent a pixel in an image, resulting in less memory used 
to store or even transmit an image.  Thus, the Karhunen-Lo\`{e}ve theorem 
enables us to pick the best 
basis thus to minimize the entropy and error, to better represent an image for 
optimal storage or transmission.  Here, \textit{optimal} means it uses least memory 
space to represent the data. i.e., instead of using 16 bits, it uses 11 bits.  So, 
the best basis found would allow us to better represent the digital image with
less storage memory.

In our next section we give the general context and definitions from operators in Hilbert space which we shall need: We discuss the particular orthonomal bases (ONBs) and frames which we use, and we recall the operator theoretic context of the Karhunen-Lo\`{e}ve theorem \cite{Ash90}.  In approximation problems involving a stochastic component (for example noise removal in time-series or data resulting from image processing) one typically ends up with correlation kernels; in some cases as frame kernels; see details in section \ref{sec:5}. In some cases they arise from systems of vectors in Hilbert space which form frames (see Definition \ref{D:frame}). In some cases parts of the frame vectors fuse (fusion frames) onto closed subspaces, and we will be working with the corresponding family of (orthogonal) projections. Either way, we arrive at a family of selfadjoint positive semidefinite operators in Hilbert space. The particular Hilbert space depends on the application at hand. While the Spectral Theorem does allow us to diagonalize these operators, the direct application the Spectral Theorem may lead to continuous spectrum which is not directly useful in computations, or it may not be computable by recursive algorithms. As a result we introduce in Section \ref{sec:7} a weighting of the operator to be analyzed.

The questions we address are optimality of approximation in a variety of ONBs, and the choice of the ``best" ONB. Here ``best" is given two precise meanings: (1) In the computation of a sequence of approximations to the frame vectors, the error terms must be smallest possible; and similarly (2) we wish to minimize the corresponding sequence of entropy numbers (referring to von Neumann's entropy). In two theorems we make precise an operator theoretic Karhunen-Lo\`{e}ve basis, which we show is optimal both in regards to criteria (1) and (2). But before we prove our theorems, we give the two problems an operator theoretic formulation; and in fact our theorems are stated in this operator theoretic context.

In section \ref{sec:7}, we introduce the weighting, and we address a third optimality criteria; that of optimal weights: Among all the choices of weights (taking the form of certain discrete probability distributions) turning the initially given operator into trace-class, the problem is then to select the particular weights which are optimal in a sense which we define precisely.

\section{General Background}
\label{sec:2}


\subsection{From Data to Hilbert Space}
\label{sec:0}
In computing probabilities and entropy Hilbert space serves as a helpful tool.  As an example take a unit vector $f$ in some fixed Hilbert space $\mathcal{H}$,
and an orthonormal basis (ONB) $\psi_{i}$ with $i$ running over an index set 
$I$. With this we now introduce two families of probability measures, one 
family $P_{f}(\cdot)$ indexed by $f \in \mathcal{H}$, and a second family 
$P_{T}$ indexed by a class of operators $T: \mathcal{H} \to \mathcal{H}$.
 
\subsubsection{The measures $P_{f}$}  Define 
\begin{equation}
\label{E:P_f}
P_{f}(A) = \sum_{i \in A}|\langle \psi_{i} | f \rangle|^{2}
\end{equation}
where $A \subset I$, and where $\langle \cdot | \cdot \rangle$ denotes the 
inner product.  Following physics convention, we make our inner product linear 
in the second variable. That will also let us make use of Dirac's convenient 
notation for rank-one operators, see eq (\ref{E:dirac2}) below.

Note then that $P_{f}(A)$ is a probability measure on the 
finite subsets $A$ of $I$.  To begin with, we make the restriction to finite 
subsets. This is merely for later use in recursive systems, see e.g., 
eq (\ref{E:P_f}). In diverse contexts, extensions from finite to infinite is 
then done by means of Kolmogorov's consistency principle \cite{Kol77}.

By introducing a weighting we show that this 
assigment also works for more general vector configurations $\mathcal{C}$ than 
ONBs.  Vectors in 
$\mathcal{C}$ may represent signals or image fragments/blocks.  Correlations 
would then be measured as inner products $\langle u | v \rangle$ with $u$ and 
$v$ representing different image pixels. Or in the case of signals, $u$ and $v$
might represent different frequency subbands.  

\subsubsection{The measures $P_{T}$} A second more general family of 
probability measures arising in the context of 
Hilbert space is called determinantal measures.  Specifically, consider 
bitstreams as points in an infinite Cartesian product 
$\Omega =  \prod_{i\in \mathbb{N}}\{0,1\}$.  Cylinder sets in $\Omega$ are 
indexed by finite subsets $A \subset \mathbb{N}$,
\[
C_{A} = \{(\omega_{1}, \omega_{2},...)| \omega_{i}=1 \quad \text{for } i \in A\}
\]
If $T$ is an operator in $\mathit{l}^{2}(\mathbb{N})$ such that 
$0 \leq  \langle u|Tu \rangle  \leq \|u\|^{2}$ for all $u \in \mathit{l}^{2}$,
then set 
\begin{equation}
\label{E:P_f}
P_{T}(C_{A}) = det(T(i,j))_{i,j \in A}
\end{equation}
where $T(i,j)$ is the matrix representation of $T$ computed in some ONB in 
$\mathit{l}^{2}$.  Using general principles \cite{Kol77, Jor06} it can be 
checked that $P_{T}(C_{A})$ is independent of the choice of ONB.

To verify that $P_{T}(\cdot)$ extends to a probability measure defined on the 
sigma-algebra generated by $C_{A}s$, see e.g. \cite{Jor06}, Ch7.  The argument is
based on Kolmogorov's consistency principle, see \cite{Kol77}

Frames (Definition \ref{D:G-op}) are popular in analyzing signals and images.  
This fact raises questions of comparing two approximations: one using a frame the 
other using an ONB.  However, there are several possible choices of ONBs.  
An especially natural choice of an ONB would be one diagonalizes the matrix 
$(\langle f_{i}|f_{j} \rangle)$ where $(f_{i})$ is a frame.  We call such a 
choice of ONB Kahunen-Lo\`{e}ve (K-L) expansion.  Section \ref{sec:KL} deals
with a continuous version of this matrix problem. The justification for why 
diagonalization occurs and works also when the frame $(f_{i})$ is infinite 
that is 
based on the Spectral Theorem. For the details regarding this, see the proof 
of Theorem \ref{T:3.3} below.

In symbols, we designate the K-L ONB associated to the frame $(f_{i})$ as
$(\phi_{i})$.  In computations, we must rely on finite sums, 
and we are interested in estimating the errors when different approximations 
are used, and where summations are truncated.  Our main results make precise 
how the K-L ONB yields better approximations, smaller entropy and better 
synthesis.  Even more we show that infimum calculations yield minimum numbers 
attained at the K-L ONB expansions. We emphasize that the ONB depends on the frame chosen, and this point will be discussed in detail later.

If larger systems are subdivided the smaller parts may be represented by
projections $P_{i}$, and the $i-j$ correlations by the operators $P_{i}P_{j}$.
The entire family $(P_{i})$ is to be treated as a fusion frame 
\cite{CaKu04, CaKuLi06}.  Fusion frames are defined in Definition 
\ref{D:fusionf} 
below. Frames themselves are generalized bases with redundancy, for example 
occurring in signal processing involving multiplexing. The fusion frames allow 
decompositions with closed subspaces as opposed to individual vectors. They 
allow decompositions of signal/image processing tasks with degrees of 
homogeneity.

\subsection{Definitions}
\label{sec:3}
\begin{definition}
\label{D:H}
Let $\mathcal{H}$ be a Hilbert space.
Let $(\psi_{i})$ and $(\phi_{i})$ be orthonormal bases (ONB), with index set $I$. 
Usually 
\begin{equation}
\label{E:index}
I= \mathbb{N}=\{1,2,...\}.
\end{equation}
If $(\psi_{i})_{i \in I}$ is an ONB, we set $Q_{n}:=$ the orthogonal projection 
onto $span\{\psi_{1}, ... , \psi_{n}\}$.
\end{definition}

We now introduce a few facts about operators which will be needed in the paper. In particular we recall Dirac's terminology \cite{Dir47} for rank-one operators in Hilbert space. While there are alternative notation available, Dirac's bra-ket terminology is especially efficient for our present considerations.
\begin{definition}
\label{D:Dirac}
Let vectors $u$, $v \in \mathcal{H}$.  Then  
\begin{equation}
\label{E:dirac1}
  \langle u | v \rangle = \text{inner product} \in \mathbb{C},
\end{equation}

\begin{equation}
\label{E:dirac2}
  |u \rangle \langle v| = \text{rank-one operator}, 
  \mathcal{H} \to \mathcal{H},
\end{equation}
 where the operator $|u \rangle \langle v|$ acts as follows

\begin{equation}
\label{E:dirac3}
|u \rangle \langle v|w = |u\rangle \langle v|w \rangle 
  = \langle v | w \rangle u, \quad \textit{for all } w \in \mathcal{H}.
\end{equation}
\end{definition}

Dirac's bra-ket and ket-bra notation is is popular in physics, and it is 
especially convenient in working with rank-one operators and inner products. 
For example, in the middle term in eq (\ref{E:dirac3}), the vector $u$ is 
multiplied by a scalar, the inner product; and the inner product comes about 
by just merging the two vectors.


\subsection{Facts}
\label{sec:4}
The following formulas reveal the simple rules for the algebra of rank-one operators, their composition, and their adjoints.
\begin{equation}
\label{E:fact1}
 |u_{1} \rangle \langle v_{1}||u_{2} \rangle \langle v_{2}| 
           = \langle v_{1} | u_{2} \rangle |u_{1}\rangle \langle v_{2}|,
\end{equation}
and
\begin{equation}
\label{E:fact2}
 |u \rangle \langle v|^{*}=|v \rangle \langle u|.
\end{equation}

In particular, formula (\ref{E:fact1}) shows that the product of two rank-one
operators is again rank-one.  The inner product $\langle v_{1}|u_{2} \rangle$ is
a measure of a correlation between the two operators on the LHS of 
(\ref{E:fact1}).

If $S$ and $T$ are bounded operators in $\mathcal{H}$, in $B(\mathcal{H})$, 
then 
\begin{equation}
\label{E:fact3}
S|u \rangle \langle v|T = |Su \rangle \langle T^{*}v|
\end{equation}

If $(\psi_{i})_{i \in \mathbf{N}}$ is an ONB then the projection 
\[
  Q_{n}:= proj span\{\psi_{1}, ... , \psi_{n}\}
\]
is given by
\begin{equation}
\label{E:fact4}
  Q_{n} = \sum_{i=1}^{n}|\psi_{i} \rangle \langle \psi_{i}|; 
\end{equation}
and for each $i$, $|\psi_{i} \rangle \langle \psi_{i}|$ is the projection onto 
the one-dimensional subspace $\mathbf{C} \psi_{i} \subset \mathcal{H}$.

\section{The Kahunen-Lo\`{e}ve transform}
\label{sec:KL}
In general, one refers to a \textit{Karhunen-Lo\`{e}ve transform} as an 
expansion in Hilbert 
space with respect to an ONB resulting from an application of the Spectral-Theorem.

\begin{example}
\label{Ex:KL}
Suppose $X_{t}$ is a stochastic process indexed by $t$ in a finite interval 
$J$, and taking values in $L^{2}(\Omega, P)$ for some probability space 
$(\Omega, P)$. Assume the normalization $E(X_{t})=0$. Suppose the integral 
kernel $E(X_{t} X_{s})$ can be 
diagonalized, i.e., suppose that 
\[
  \int_{J}{E(X_{t} X_{s})\phi_{k}(s)}ds=\lambda_{k}\phi_{k}(t)
\]
with an ONB $(\phi_{k})$ in $L^{2}(J)$.  If $E(X_{t})=0$ then
\[
  X_{t}(\omega)=\sum_{k}\sqrt{\lambda_{k}}\phi_{k}(t)Z_{k}(\omega), 
  \quad \omega \in \Omega
\]
where $E(Z_{j} Z_{k})=\delta_{j,k}$, and $E(Z_{k})=0$.
The ONB $(\phi_{k})$ is called the \textit{KL-basis} with respect to the 
stochastic processes $\{X_{t}: t \in I \}$.

The KL-theorem \cite{Ash90} states that if $(X_{t})$ is Gaussian, then so are 
the random variables $(Z_{k})$.  Furthermore, they are $N(0,1)$ i.e., normal
with mean zero and variance one, so independent and identically distributed. 
This last fact explains the familiar \textit{optimality} of KL in transform 
coding.
\end{example}

\begin{remark}
\label{R:KL}

Consider the case when 
\[
  E(X_{t} X_{s})=\frac{1}{2}(t^{2H}+s^{2H}-|t-s|^{2H})
\]
and $H \in (0,1)$ is fixed. If $J= \mathbb{R}$ in the above application of KL 
to stochastic processes then it is possible by a fractional 
integration to make the $L^{2}(\mathbb{R})$-ONB consist of wavelets, i.e.,
\[
  \psi_{j,k}(t):=2^{j/2}\psi(2^{j}t-k), \quad j,k \in \mathbb{Z}, 
\text{ i.e. double-indexed,} 
 \quad t\in \mathbb{R}, \text{ for some } \psi \in L^{2}(\mathbb{R})
\]
see e.g. \cite{Jor06}.
The process $X_{t}$ is called $H-$fractional Brownian motion, as outlined in 
e.g. \cite{Jor06} p.57.
\end{remark}

The following theorem makes clear the connection to Hilbert space geometry as 
used in present paper:

\begin{theorem}
\label{T:3.3}
Let $(\Omega, P)$ by a probability space, $J \subset \mathbb{R}$ an interval 
(possibly infinite), and let $(X_{t})_{t\in J}$ be a stochastic process with 
values in $L^{2}(\Omega, P)$.  Assume $E(X_{t})=0$ for all $t \in J$.  Then
$L^{2}(J)$ splits as an orthogonal sum 
\begin{equation}
\label{E:Lsum}
  L^{2}(J)=\mathcal{H}_{d}\oplus \mathcal{H}_{c}
\end{equation} 
(d is for discrete and c is for continuous) such that the following data 
exists:
\begin{itemize}
  \item[(a)] $(\phi_{k})_{k \in \mathbb{N}}$ an ONB in $\mathcal{H}_{d}$.
  \item[(b)] $(Z_{k})_{k \in \mathbb{N}}$ : independent random variables.
  \item[(c)] $E(Z_{j} Z_{k})=\delta_{j,k}$, and $E(Z_{k})=0$.
  \item[(d)] $(\lambda_{k}) \subset \mathbb{R}_{\geq 0}$.
  \item[(e)] $\phi(\cdot,\cdot)$ : a Borel measure on $\mathbb{R}$ in the first 
variable, such that 
  \begin{itemize}
    \item[(i)] $\phi(E, \cdot) \in \mathcal{H}_{c}$ for $E$ an open 
subinterval of $J$.
  \end{itemize} and
  \begin{itemize}
    \item[(ii)] $\langle\phi(E_{1}, \cdot)| \phi(E_{2}, \cdot)
    \rangle_{L^{2}(J)} =0$ whenever $E_{1} \cap E_{2} = \emptyset$.
  \end{itemize}
  \item[(f)] $Z(\cdot, \cdot)$ : a measurable family of random variables such
  that $Z(E_{1}, \cdot)$ and $Z(E_{2}, \cdot)$ are independent when
  $E_{1}, E_{2} \in \mathcal{B}_{J}$ and $E_{1} \cap E_{2} = \emptyset$, 
  \[
    E(Z(\lambda, \cdot) Z(\lambda', \cdot))=\delta(\lambda-\lambda'), 
    \text{ and } E(Z(\lambda, \cdot))=0.
  \]
\end{itemize}
Finally, we get the following Karhunen-Lo\`{e}ve expansions for the 
$L^{2}(J)$-operator with integral kernel $E(X_{t} X_{s})$:
\begin{equation}
\label{E:Suml}
  \sum_{k \in \mathbb{N}} \lambda_{k}|\phi_{k} \rangle \langle \phi_{k}|
  + \int_{J}{\lambda|\phi(d \lambda, \cdot)\rangle \langle 
    \phi(d \lambda, \cdot)|}
\end{equation}
Moreover, the process decomposes thus:
\begin{equation}
\label{E:Sumsqrtl}
  X_{t}(\omega)= \sum_{k\in \mathbb{N}}\sqrt{\lambda_{k}}Z_{k}(\omega)
  \phi_{k}(t)+\int_{J}{\sqrt{\lambda}Z(\lambda, \omega)\phi(d\lambda, t)}. 
\end{equation}
\end{theorem}
\begin{proof}
By assumption the integral operator in $L^{2}(J)$ with kernel $E(X_{t} X_{s})$ 
is selfadjoint, positive semidefinite, but possibly unbounded. By the Spectral
Theorem, this operator has the following representation.
\[
  \int_{0}^{\infty}\lambda Q(d\lambda) 
\]
where $Q(\cdot)$ is a projection valued measure defined on the Borel subsets
$\mathcal{B}$, of $\mathbb{R}_{\geq 0}$.  Recall
\[
  Q(S_{1} \cap S_{2}) =  Q(S_{1})Q(S_{2}) \quad \text{for } 
  S_{1}, S_{2} \in \mathcal{B};
\]
and $\int_{0}^{\infty}Q(d\lambda)$ is the identity operator in $L^{2}(J)$.  The
two closed subspaces $\mathcal{H}_{d}$ and $\mathcal{H}_{c}$ in the 
decomposition (\ref{E:Lsum}) are the discrete and continuous parts of the projection
value measure $Q$, i.e., $Q$ is discrete (or atomic) on $\mathcal{H}_{d}$, and
it is continuous on $\mathcal{H}_{c}$.

Consider first
\[
  Q_{d}(\cdot)=Q(\cdot)|_{\mathcal{H}_{d}} 
\]
and let $(\lambda_{k})$ be the atoms.  Then for each $k$, the non-zero 
projection $Q(\{\lambda_{k}\})$ is a sum of rank one projections 
$|\phi_{k} \rangle \langle \phi_{k}|$ corresponding to a choice of ONB in the
$\lambda_{k}$ subspace. (Usually the multiplicity is one, in which case 
$Q(\{\lambda_{k}\})=|\phi_{k} \rangle \langle \phi_{k}|$.)
This accounts for the first terms in the representations (\ref{E:Suml}) 
and (\ref{E:Sumsqrtl}).  

We now turn to the continuous part, i.e., the subspace $\mathcal{H}_{c}$, and 
the continuous projection valued measure
\[
  Q_{c}(\cdot)=Q(\cdot)|_{\mathcal{H}_{c}}. 
\]
The second terms in the two formulas (\ref{E:Suml}) and (\ref{E:Sumsqrtl}) 
result from an application of a disintegration theorem from \cite{DuJo07}, 
Theorem 3.4. This theorem is applied to the measure $Q_{c}(\cdot)$.

We remark for clarity that the term 
$|\phi(d \lambda, \cdot)\rangle \langle \phi(d \lambda, \cdot)|$ 
under the integral sign in (\ref{E:Suml}) is merely a measurable field of 
projections $P(d\lambda)$.
\end{proof}

Our adaptation of the spectral theorem from books in operator theory 
(e.g., \cite{Jor06}) is made with view to the application at hand, and our 
version of Theorem \ref{T:3.3} serves to make the adaptation to how operator 
theory is used for time series, and for encoding. We have included it here 
because it isn't written precisely this way elsewhere.

\section{Frame Bounds and Subspaces}
\label{sec:5}
The word `frame' in the title refers to a family of vectors in Hilbert space 
with basis-like properties which are made precise in Definition \ref{D:frame}.
We will be using entropy and information as defined classically by Shannon 
\cite{Sh60}, and extended to operators by von Neumann \cite{GvN48}.

The reference \cite{Ash90} offers a good overview of the basics of both. 
Shannon's pioneering idea was to quantify digital ``information," essentially 
as the negative of entropy, entropy being a measure of ``disorder." This idea 
has found a variety of application n both signal/image processing, and in 
quantum information theory, see e.g., \cite{Kr05}. A further recent use of 
entropy is in digital encoding of signals and images, compressing and 
quantizing digital information into a finite floating-point computer register. 
(Here we use the word ``quantizing" \cite{SkChEb01}, \cite{Smi02}, \cite{Use01} 
in the sense of computer science.) To 
compress data for storage, an encoding is used which takes into consideration 
probability of occurrences of the components to be quantized; and hence 
entropy is a gauge for the encoding. 

\begin{definition}
\label{D:traceclass}
$T \in B(\mathcal{H})$ is said to be trace class if and only if 
$\sum \langle \psi_{i}| T\psi_{i} \rangle$ is absolutely convergent for some 
ONB $(\psi_{i})$.  In this case, set 
\begin{equation}
\label{E:trace}
tr(T):= \sum_{i}^{n} \langle \psi_{i}|T\psi_{i}\rangle
\end{equation}
\end{definition}

\begin{definition}
\label{D:frame}
A sequence $(h_{\alpha})_{\alpha \in A}$ in $\mathcal{H}$ is called a frame if 
there are constants $0<c_{1} \leq c_{2} < \infty$ such that

\begin{equation}
\label{E:framebd}
c_{1}\|f\|^{2} \leq \sum_{\alpha \in A}|\langle h_{\alpha}|f\rangle|^{2} \leq 
c_{2}\|f\|^{2}
\text{ for all } f \in \mathcal{H}.
\end{equation}
\end{definition}

\begin{definition}
\label{D:G-op}
Suppose we are given a frame operator
\begin{equation}
\label{E:g_op}
G=\sum_{\alpha \in A}w_{\alpha}|f_{\alpha} \rangle \langle f_{\alpha}|
\end{equation}
and an ONB $(\psi_{i})$.  Then for each $n$, the numbers
\begin{equation}
\label{E:errorterm}
E_{n}^{\psi}= \sum_{\alpha \in A} w_{\alpha} \|f_{\alpha} - \sum_{i=1}^{n}\langle \psi_{i}|f_{\alpha}\rangle \psi_{i}\|^{2}
\end{equation}
are called the error-terms.
\end{definition}

Set $L:\mathcal{H} \to \mathit{l}^{2}$, 
\begin{equation}
\label{E:map_L}
L:f \mapsto (\langle h_{\alpha}|f\rangle)_{\alpha \in A}.
\end{equation}

\begin{lemma}
\label{L:L}
If $L$ is as in (\ref{E:map_L}) then $L^{*}: \mathit{l}^{2} \to \mathcal{H}$ is
given by
\begin{equation}
\label{E:L*}
L^{*}((c_{\alpha}))=\sum_{\alpha \in A}c_{\alpha}h_{\alpha}
\end{equation}
where $(c_{\alpha})\in \mathit{l}^{2}$; and
\begin{equation}
\label{E:L*L}
L^{*}L=\sum_{\alpha \in A} |h_{\alpha} \rangle \langle h_{\alpha}|
\end{equation}
\end{lemma}

\begin{lemma}
\label{L:L}
If $(f_{\alpha})$ are the normalized vectors resulting from a frame 
$(h_{\alpha})$, i.e., $h_{\alpha}= \|h_{\alpha}f_{\alpha}\|$, and 
$w_{\alpha}:= \|h_{\alpha}\|^{2}$, then $L^{*}L$ has the form 
(\ref{E:frame_operator}).
\end{lemma}

\begin{proof}
The desired conclusion follows from the Dirac formulas 
(\ref{E:fact1})-(\ref{E:fact2}).  Indeed 
\[
|h_{\alpha} \rangle \langle h_{\alpha}|
=|\|h_{\alpha}\|f_{\alpha}\rangle \langle\|h_{\alpha}\|f_{\alpha}|
=\|h_{\alpha}\|^{2}|f_{\alpha}\rangle \langle f_{\alpha}|=w_{\alpha}P_{\alpha}
\]
where $P_{\alpha}$ satisfies the two rules 
$P_{\alpha}=P_{\alpha}^{*}=P_{\alpha}^{2}$.
\end{proof}

\begin{definition}
\label{D:frameop}
Suppose we are given $(f_{\alpha})_{\alpha \in A}$, a frame, non-negative numbers  
$\{w_{\alpha}\}_{\alpha \in A}$, where $A$ is an index set, with 
$\|f_{\alpha}\|=1$, for all $\alpha \in A$.
\begin{equation}
\label{E:frame_operator}
G:= \sum_{\alpha \in A} w_{\alpha} |f_{\alpha} \rangle \langle f_{\alpha}|
\end{equation}
is called a \textbf{frame} operator associated to $(f_{\alpha})$.
\end{definition}

\begin{lemma}
\label{L:trG}
Note that $G$ is trace class if and only if $\sum_{\alpha}w_{\alpha} < \infty$; and then
\begin{equation}
\label{E:trG}
trG=\sum_{\alpha \in A}w_{\alpha}
\end{equation}
\end{lemma}

\begin{proof}
The identity (\ref{E:trG}) follows from the fact that all the rank-one operators
$|u \rangle \langle v|$ are trace class, with 
\[tr|u \rangle \langle v|=\sum_{i=1}^{n}\langle \psi_{i}|u\rangle \langle v|\psi_{i}\rangle = \langle u | v \rangle.\]
In particular, $tr|u\rangle \langle u| = \|u\|^{2}$.
\end{proof}

We shall consider more general frame operators
\begin{equation}
\label{E:gen_fram_op}
G=\sum_{\alpha \in A} w_{\alpha}P_{\alpha}
\end{equation}
where $(P_{\alpha})$ is an indexed family of projections in $\mathcal{H}$, ie., 
$P_{\alpha}=P_{\alpha}^{*}=P_{\alpha}^{2}$, for all $\alpha \in A$.  Note that 
$P_{\alpha}$ is trace class if and only if it is finite-dimensional, ie., if and
only if the subspace $P_{\alpha}\mathcal{H}=\{x \in \mathcal{H}|P_{\alpha}x=x\}$ is finite-dimensional.

When $(\psi_{i})$ is given set $Q_{n}:=\sum_{i=1}^{n}|\psi_{i} \rangle \langle \psi_{i}|$ and $Q_{n}^{\bot}=I-Q_{n}$ where $I$ is the identity operator in $\mathcal{H}$.

\begin{lemma}
\label{L:error}
\begin{equation}
\label{E:error}
E_{n}^{\psi}=tr(GQ_{n}^{\bot}).
\end{equation}
\end{lemma}
\begin{proof}
The proof follows from the previous facts, using that
\[
\|f_{\alpha}-Q_{n}f_{\alpha}\|^{2}=\|f_{\alpha}\|^{2}-\|Q_{n}f_{\alpha}\|^{2}
\]
for all $\alpha \in A$ and $n \in \mathbf{N}$. The expression \ref{E:errorterm} for the error term is motivated as follows.  The vector components $f_{\alpha}$
in Definition \ref{E:frame_operator} are indexed by $\alpha \in A$ and are assigned weights $w_{\alpha}$. But rather than computing $\sum_{\alpha}w_{\alpha}$ as
in Lemma \ref{L:trG}, we wish to replace the vectors $f_{\alpha}$ with finite
approximations $Q_{n}f_{\alpha}$, then the error term \ref{E:errorterm} measures
how well the approximation fits the data.  

\begin{lemma}
\label{L:trGQ_n}
  $tr(GQ_{n})=\sum_{\alpha \in A} w_{\alpha}\|Q_{n}f_{\alpha}\|^{2}$.
\end{lemma}
\begin{proof}
  \[
    tr(GQ_{n})= \sum_{i} \langle \psi_{i} | GQ_{n}\psi_{i} \rangle 
    = \sum_{i} \langle Q_{n}\psi_{i} | GQ_{n}\psi_{i} \rangle 
    = \sum_{i=1}^{n} \langle \psi_{i} | G\psi_{i} \rangle 
  \]
  \[
    = \sum_{\alpha \in A}w_{\alpha} \sum_{i=1}^{n}|\langle \psi_{i} | 
      f_{\alpha} \rangle|^{2} 
    = \sum_{\alpha \in A} w_{\alpha}\|Q_{n}f_{\alpha}\|^{2}, 
  \] as claimed. 
\end{proof}

\textit{Proof of Lemma \ref{L:error} continued}. The relative error is represented by the difference:
\[
  \sum_{\alpha \in A} w_{\alpha} - 
     \sum_{\alpha \in A} w_{\alpha}\|Q_{n}f_{\alpha}\|^{2} 
  = \sum_{\alpha \in A} w_{\alpha}\|f_{\alpha}\|^{2} -
     \sum_{\alpha \in A} w_{\alpha}\|Q_{n}f_{\alpha}\|^{2} 
\]
\[
  = \sum_{\alpha \in A} w_{\alpha}(\|f_{\alpha}\|^{2} - 
    \|Q_{n}f_{\alpha}\|^{2}) 
  = \sum_{\alpha \in A} w_{\alpha}\|f_{\alpha} - Q_{n}f_{\alpha}\|^{2}
\]
\[
  = \sum_{\alpha \in A} w_{\alpha}\|Q_{n}^{\perp}f_{\alpha}\|^{2}
  = tr(GQ_{n}^{\perp}).
\]
\end{proof}

\begin{definition}
\label{D:error_seq}
If $G$ is a more general frame operator (\ref{E:gen_fram_op}) and $(\psi_{i})$ is some 
ONB, we shall set $E_{n}^{\psi}:=tr(GQ_{n}^{\bot})$; this is called the error 
sequence.
\end{definition}

The more general case of (\ref{E:gen_fram_op}) where
\begin{equation}
\label{E:rankP}
rank P_{\alpha} = trP_{\alpha}>1
\end{equation}
corresponds to what are called subspace frames, i.e., indexed families
$(P_{\alpha})$ of orthogonal projections such that there are 
$0<c_{1} \leq c_{2} < \infty$ and weights $w_{\alpha} \geq 0$ such that
\begin{equation}
\label{E:cPf}
c_{1}\|f\|^{2} \leq \sum_{\alpha \in A}w_{\alpha}\|P_{\alpha}f\|^{2} \leq 
c_{2}\|f\|^{2}
\end{equation}
for all $f \in \mathcal{H}$.

We now make these notions precise:

\begin{definition} 
\label{D:}
A \textbf{projection} in a Hilbert space $\mathcal{H}$ 
is an operator $P$ in $\mathcal{H}$ satisfying $P^{*} = P = P^{2}$. It is 
understood that our projections $P$ are orthogonal, i.e., that $P$ is a 
\textbf{selfadjoint} idempotent. The orthogonality is essential because, 
by von Neumann, we know that there is then a 1-1 correspondence between 
closed subspaces in $\mathcal{H}$ and (orthogonal) projections: every 
closed subspace in $\mathcal{H}$ is the range of a unique projection. 
\end{definition}

We shall need the following generalization of the notion Definition 
\ref{D:frame} of frame.

\begin{definition} 
\label{D:fusionf}   
A \textbf{fusion frame} (or subspace frame) in a Hilbert space $\mathcal{H}$ 
is an indexed system ($P_{i}$ , $w_{i}$) where each $P_{i}$  is a 
projection, and where ($w_{i}$)  is a system of numerical weights, 
i.e., $w_{i}  >  0$,  such that (\ref{E:cPf}) holds: specifically, 
the system ($P_{i}$ , $w_{i}$) is called a fusion frame when 
(\ref{E:cPf}) holds. 
\end{definition}

It is clear (see also section 6) that the notion of ``fusion frame" 
contains conventional frames Definition \ref{D:frame} as a special case.

The property (\ref{E:cPf}) for a given system controls the weighted 
overlaps of the variety of subspaces  $\mathcal{H}_{i} (:= P_{i}(\mathcal{H}))$ 
making up the system, i.e., the intersections of subspaces corresponding 
to different values of the index. Typically the pair wise intersections are 
non-zero. The case of zero pair wise intersections happens precisely when 
the projections are orthogonal, i.e., when  $P_{i} P_{j} = 0$ for all pairs 
with $i$ and $j$  different. In frequency analysis, this might represent 
orthogonal frequency bands.

When vectors in $\mathcal{H}$ represent signals, we think of bands of signals 
being ``fused" into the individual subspaces $\mathcal{H}_{i}$. Further, note that 
for a given system of subspaces, or equivalently, projections, there may 
be many choices of weights consistent with (\ref{E:cPf}): The overlaps 
may be controlled, or weighted, in a variety of ways. The choice of weights 
depends on the particular application at hand.

\begin{theorem}
\label{T:smallest_error}
The Karhunen-Lo\`{e}ve ONB with respect to the frame operator $L^{*}L$ gives 
the smallest error terms in the approximation to a frame operator.
\end{theorem}
\begin{proof}
Given the operator $G$ which is trace class and positive semidefinite, we may 
apply the spectral theorem to it. What results is a discrete spectrum, with the
natural order $\lambda_{1} \geq \lambda_{2} \geq ...$ and a corresponding ONB 
$(\phi_{k})$ consisting of eigenvectors, i.e., 
\begin{equation}
\label{E:eigenvector}
G\phi_{k}=\lambda_{k}\phi_{k}, k \in \mathbb{N}
\end{equation}
called the Karhunen-Lo\`{e}ve data.  The spectral data may be constructed 
recursively starting with 
\begin{equation}
\label{E:lambda1}
\lambda_{1}=\underset{\phi \in \mathcal{H}, \|\phi\|=1}{sup}
\langle \phi| G\phi \rangle 
= \langle \phi_{1}| G\phi_{1} \rangle
\end{equation}
and
\begin{equation}
\label{E:lambdak+1}
\lambda_{k+1}=\underset{\underset{\phi \bot \phi_{1}, \phi_{2}, ..., \phi_{k}}
{\phi \in \mathcal{H}, \|\phi\|=1}}{sup}
\langle \phi| G\phi \rangle 
= \langle \phi_{k+1}| G\phi_{k+1} \rangle
\end{equation}
Now an application of \cite{ArKa06}; Theorem 4.1 yields 
\begin{equation}
\label{E:ineq}
\sum_{k=1}^{n}\lambda_{k} \geq tr(Q_{n}^{\psi}G) 
  = \sum_{k=1}^{n}\langle \psi_{k}| G\psi_{k} \rangle \quad \text{for all }n,
\end{equation} 
where $Q_{n}^{\psi}$ is the sequence of projections from (\ref{E:fact4}), 
deriving from some ONB $(\psi_{i})$ and arranged such that
\begin{equation}
\label{E:psi_ineq}
\langle \psi_{1} | G\psi_{1} \rangle \geq \langle \psi_{2} | G\psi_{2} \rangle 
\geq ... \quad \text{.} 
\end{equation}
Hence we are comparing ordered sequences of eigenvalues with sequences of 
diagonal matrix entries.

Finally, we have
\[
trG=\sum_{k=1}^{\infty} \lambda_{k} 
=\sum_{k=1}^{\infty} \langle \psi_{k} | G\psi_{k} \rangle < \infty.
\]


The assertion in Theorem \ref{T:smallest_error} is the validity of
\begin{equation}
\label{E:error_ineq}
E_{n}^{\phi} \leq E_{n}^{\psi}
\end{equation}
for all $(\psi_{i}) \in ONB(\mathcal{H})$, and all $n=1,2,...$; and moreover, 
that the infimum on the RHS in (\ref{E:error_ineq}) is attained for the KL-ONB 
$(\phi_{k})$. But in view of our lemma for $E_{n}^{\psi}$, Lemma \ref{L:error} 
we see that (\ref{E:error_ineq}) is equivalent to the system (\ref{E:ineq}) in 
the Arveson-Kadison theorem.
\end{proof}

The Arveson-Kadison theorem is the assertion (\ref{E:ineq}) for trace class operators, see e.g., refs \cite{Arv06} and \cite{ArKa06}. That (\ref{E:error_ineq}) is equivalent to (\ref{E:ineq}) follows from the definitions. 

\begin{remark}
Even when the operator $G$ is not trace class, there is still a conclusion 
about the relative error estimates.  With two $(\phi_{i})$ and $(\psi_{i})$ 
$\in$ 
$ONB(\mathcal{H})$ and with $n < m$, $m$ large, we may introduce the following 
relative error terms:
\[
  E_{n,m}^{\phi} = tr(G(Q_{m}^{\phi}-Q_{n}^{\phi}))
\]
and
\[
  E_{n,m}^{\psi} = tr(G(Q_{m}^{\psi}-Q_{n}^{\psi})).
\]
If $m$ is fixed, we then choose a Karhunen-Lo\`{e}ve basis $(\phi_{i})$ for 
$Q_{m}^{\psi}GQ_{m}^{\psi}$ and the following error inquality holds:
\[
  E_{n,m}^{\phi, KL} \leq E_{n,m}^{\psi}.
\]
\end{remark}

Our next theorem gives Karhunen-Lo\`{e}ve optimality for sequences of entropy
numbers.
\begin{theorem}
\label{T:smallest_entropy}
The Karhunen-Lo\`{e}ve ONB gives the smallest sequence of entropy numbers in the approximation to a frame operator.
\end{theorem}
\begin{proof}
We begin by a few facts about entropy of trace-class operators $G$.  The entropy
is defined as
\begin{equation}
\label{E:entropy}
S(G):= -tr(G\log{G}).
\end{equation}
The formula will be used on cut-down versions of an initial operator $G$.  In 
some cases only the cut-down might be trace-class.
Since the Spectral Theorem applies to $G$, the RHS in (\ref{E:entropy}) is also
\begin{equation}
\label{E:entropy_sum}
S(G)=-\sum_{k=1}^{\infty}\lambda_{k}\log{\lambda_{k}}.
\end{equation}
For simplicity we normalize such that $1=trG=\sum_{k=1}^{\infty}\lambda_{k}$,
and we introduce the partial sums
\begin{equation}
\label{E:entropy_part_sum1}
S_{n}^{KL}(G):=-\sum_{k=1}^{n}\lambda_{k}\log{\lambda_{k}}.
\end{equation}
and
\begin{equation}
\label{E:entropy_part_sum2}
S_{n}^{\psi}(G):=-\sum_{k=1}^{n}\langle \psi_{k}|G\psi_{k}\rangle 
\log{\langle \psi_{k}|G\psi_{k}\rangle}.
\end{equation}

Let $(\psi_{i}) \in ONB(\mathcal{H})$, and set 
$d_{k}^{\psi}:=\langle \psi_{k}|G\psi_{k}\rangle$; then the inequalities 
(\ref{E:ineq}) take the form:
\begin{equation}
\label{E:d_lambda}
  tr(Q_{n}^{\psi}G)=\sum_{i=1}^{n}d_{i}^{\psi} \leq \sum_{i=1}^{n} \lambda_{i}, \quad n=1,2,...
\end{equation}
where as usual an ordering
\begin{equation}
\label{E:d_order}
d_{1}^{\psi} \geq d_{2}^{\psi} \geq ...
\end{equation}
has been chosen.

Now the function $\beta(t):=t\log{t}$ is convex.  And application of Remark 6.3 in
\cite{ArKa06} then yields 
\begin{equation}
\label{E:beta_ineq}
\sum_{i=1}^{n}\beta(d_{i}^{\psi}) \leq \sum_{i=1}^{n}\beta(\lambda_{i}), 
\quad n=1,2,... \quad \text{.}
\end{equation}
Since the RHS in (\ref{E:beta_ineq}) is $-tr(G\log{G})=-S_{n}^{KL}(G)$, the desired
inequalities
\begin{equation}
\label{E:S_ineq}
S_{n}^{KL}(G) \leq S_{n}^{\psi}(G), \quad n=1,2, ...
\end{equation}
follow. i.e., the KL-data minimizes the sequence of entropy numbers.
\end{proof}

\subsubsection{Supplement}
\label{S:Suppl}
Let $G$, $\mathcal{H}$ be as before $\mathcal{H}$ is an $\infty-$dimensional 
Hilbert Space $G=\sum_{\alpha}P_{\alpha}$, $\omega_{\alpha} \geq 0$, 
$P_{\alpha}=P_{\alpha}^{*}=P_{\alpha}^{2}$.  Suppose $dim \mathcal{H}_{\lambda_{1}}(G)>0$
where $dim \mathcal{H}_{\lambda_{1}}(G)=\{\phi \in \mathcal{H} | 
G\phi = \phi \}$ and $\lambda_{1}:= sup \{\langle f|Gf\rangle$, 
$f \in \mathcal{H}$, $\|f\|=1\}$, then define $\lambda_{2}, \lambda_{3}, ...$ 
recursively 
\[
\lambda_{k+1} := sup \{\langle f|Gf\rangle|f\perp \phi_{1},\phi_{2},...,\phi_{k}\}
\]
where $dim \mathcal{H}_{\lambda_{k}}(G)>0$.
Set $\mathcal{K}=\bigvee_{k=1}^{\infty} span\{\phi_{1},\phi_{2},...,\phi_{k}\}$. Set $\rho:=inf\{\lambda_{k}|k=1,2,...\}$ then we can apply theorems 
\ref{T:smallest_error} and \ref{T:smallest_entropy} to the restriction 
$(G-\rho I)|_{\mathcal{K}}$. i.e., the operator $\mathcal{K} \to \mathcal{K}$ 
given by $\mathcal{K} \ni x \longmapsto Gx-\rho x \in \mathcal{K}$.

Actually there are two cases for G as for $G_{\mathcal{K}}:=G-\rho I$: (1) compact, (2) trace-class.
We did (2), but we now discuss (1):

When $G$ or $G_{\mathcal{K}}$ is given, we want to consider
\[
\begin{cases}
  \rho = inf \{\langle f|Gf\rangle | \|f\|=1\} \\
  \lambda_{1} = sup\{\langle f|Gf\rangle | \|f\|=1\}
\end{cases}
\]

If $G=\sum_{\alpha}\omega_{\alpha}P_{\alpha}$ then 
$\langle f|Gf\rangle=\sum_{\alpha}\omega_{\alpha}\|P_{\alpha}f\|^{2}$.  If 
$G=\sum_{\alpha}|h_{\alpha} \rangle \langle h_{\alpha}|$ where 
$(h_{\alpha}) \subset \mathcal{H}$ is a family of vectors, then 
\[
  \langle f|Gf\rangle = \sum_{\alpha}|\langle h_{\alpha}|f \rangle|^{2}.
\]

The frame bound condition takes the form
\[
  c_{1}\|f\|^{2} \leq \sum \omega_{\alpha} \|P_{\alpha}f\|^{2} \leq c_{2}\|f\|^{2}
\]
or in the standard frame case $\{h_{\alpha}\}_{\alpha \in A} \in 
FRAME(\mathcal{H})$
\[
  \langle f|Gf\rangle = \sum_{\alpha}|\langle h_{\alpha}|f \rangle|^{2}\leq c_{2}\|f\|^{2}.
\]

\begin{lemma}
\label{L:}
If a frame system (fusion frame or standard frame) has frame bounds 
$0 < c_{1} \leq c_{2} < \infty$, then the spectrum of the operator $G$ is 
contained in the closed interval 
$[c_{1}, c_{2}]=\{x\in \mathbb{R}| c_{1} \leq x \leq c_{2}\}$.
\end{lemma}
\begin{proof}
It is clear from the formula for $G$ that $G=G^{*}$. Hence the spectrum theorem
applies, and the result follows.  In fact, if 
$z \in \mathbb{C}\setminus[c_{1}, c_{2}]$ then 
\[
  \|zf-Gf\| \geq dist(z, [c_{1}, c_{2}])\|f\|
\]
so $zI-G$ is invertible.  So $z$ is in the resolvent set. Hence 
$spec(G) \subset [c_{1}, c_{2}]$.
\end{proof}

A frame is a system of vectors which satisfies the two a priory estimates 
(\ref{E:framebd}), so by the Dirac notation, it may be thought of as a statement about rank-one projections. The notion of fusion frame is the same, only with finite-rank projections; see (\ref{E:cPf}). 

\section{Splitting off rank-one operators}
\label{sec:6}
The general principle in frame analysis is to make a recursion which introduces
rank-one operators, see Definition \ref{D:Dirac} and Facts section 
\ref{sec:4}. The theorem we will proves accomplishes that for general class of 
operators in infinite dimensional Hilbert space.  Our result may also be 
viewed as extension of Perron-Frobenius's theorem for positive matrices.  
Since we do not introduce positivity in the present section, our theorem will 
instead include assumptions which restrict the spectrum of the operators to 
which our result applies.

One way rank-one operators enter into frame analysis is through equation 
 (\ref{E:L*L}).  Under the assumptions in Lemma \ref{L:L} the operator $L^{*}L$ is
invertible.  If we multiply equation (\ref{E:L*L}) on the left and on the right 
with $(L^{*}L)^{-1}$ we arrive at the following two representations for the 
identity operator 
\begin{equation}
\label{E:htilde}
  I= \sum_{\alpha}|\tilde{h}_{\alpha} \rangle \langle h_{\alpha}| 
   = \sum_{\alpha} |h_{\alpha} \rangle \langle \tilde{h}_{\alpha}|
\end{equation}
where $\tilde{h}_{\alpha} = (L^{*}L)^{-1}h_{\alpha}$.

Truncation of the sums in (\ref{E:htilde}) yields non-selfadjoint operators 
which are used in approximation of data with frames.  Starting with a general
non-selfadjoint operator $T$, our next theorem gives a general method for 
splitting off a rank-one operator from $T$.

\begin{theorem}
\label{T:split}
Let $\mathcal{H}$ be a generally infinite dimensional Hilbert space, and let $T$
be a bounded operator in $\mathcal{H}$.  Under the following three assumptions
we can split off a rank-one operator from $T$.  Specifically assume 
$a \in \mathbb{C}$ satisfies:
\begin{itemize}
\item[(1)] $0 \ne a \in spec(T)$ where $spec(T)$ denotes the spectrum of $T$.
\item[(2)] $dim R(a-T)^{\perp} =1$ where $R(a-T)$ denotes the range of the 
operator $aI-T$, and $\perp$ denotes the orthogonal complement.
\item[(3)] $lim_{n\to \infty} a^{-n}T^{n}x=0$ for all $x \in R(a-T)$.  
\end{itemize}
Then it follows that the limit exists everywhere in $\mathcal{H}$ in the strong
topology of $\mathcal{B}(\mathcal{H})$.  Moreover, we may pick the following 
representation
\begin{equation}
\label{E:limaT}
  lim_{n \to \infty} a^{-n}T^{n} = |\xi \rangle \langle w_{1}|
\end{equation}
for the limiting operator on $\mathcal{H}$, where 
\begin{equation}
\label{E:w}
  \|w_{1}\|=1, \quad T^{*}w_{1} = \bar{a}w_{1}, \quad 
  \langle \xi | w_{1} \rangle = 1,
\end{equation}
in fact 
\begin{equation}
\label{E:xiw}
  \xi-w_{1} \in \overline{R(a-T)}, \quad and \quad T\xi = a\xi,
\end{equation}
where the over-bar denotes closure.  
\end{theorem}

Theorem \ref{T:split} is an analogue of the Perron-Frobenius theorem 
(e.g., \cite{Jor06}). Dictated by our applications,  the present 
Theorem \ref{T:split} is adapted to a different context where the present 
assumptions are different than those in the Perron-Frobenius theorem. We have 
not seen it stated in the literature in this version, and the proof (and 
conclusions) are different from that of the standard Perron-Frobenius theorem.

\begin{remark}
For the reader's benefit we include the following statement of the 
Perron-Frobenius theorem in a formulation which makes it clear how 
Theorem \ref{T:split} extends this theorem.
\subsubsection{Perron-Frobenius:} Let $d<\infty$ be given and let $T$ be a $d \times d$ matrix with entries $T_{i,j} \geq 0$, and with positive spectral radius 
$a$.  Then there is a column-vector $w$ with $w_{i} \geq 0$, and a row-vector 
$\xi$ such that the following conditions hold:
\[
Tw=aw, \quad \xi T = a\xi, \quad and \quad \xi w = 1.
\]
 
\end{remark}

\begin{proof}(of Theorem \ref{T:split})
Note that by the general operator theory, we have the following formulas:
\[
  R(a-T)^{\perp} = N(T^{*}-\bar{a})
  = \{y \in \mathcal{H}|T^{*}y=\bar{a}y\}.
\]  
By assumption (2) this is a one-dimensional space, and we may pick $w_{1}$ such
that $T^{*}w_{1}=\bar{a}w_{1}$, and $\|w_{1}\|=1$.  This means that
\[
  \{\mathbb{C}w_{1}\}^{\perp} = R(a-T)^{\perp \perp} = \overline{R(a-T)}
\]
is invariant under $T$.  

As a result, there is a second bounded operator $G$ which maps the space
$\overline{R(a-T)}$ into itself, and restricts $T$, i.e. 
$T|_{\overline{R(a-T)}}=G$.  Further, there is a vector 
$\eta^{\perp} \in (\mathbb{C}w_{1})^{\perp}$ such that $T$ has the following 
matrix representation
\[ 
  \mathbb{C}w_{1} \quad (w_{1})^{\perp}
\]
\[
  T=
  \begin{pmatrix}
    a & | & 0 0 \cdots \\
    -- & | & --- \\
    \eta^{\perp} & | & G
  \end{pmatrix}
  \begin{array}{c}
    \mathbb{C}w_{1} \\
    \\
    (w_{1})^{\perp}
  \end{array}.
\]
The entry $a$ in the top left matrix corner represents the following operator, 
$sw_{1} \mapsto asw_{1}$.  The vector $\eta_{\perp}$ is fixed, and 
$Tw_{1}= aw_{1}+\eta^{\perp}$.  The entry $\eta^{\perp}$ in the bottom left 
matrix corner represents the operator $sw_{1} \mapsto s\eta^{\perp}$, or
$|\eta^{\perp} \rangle \langle w_{1}|$.

In more detail: If $Q_{1}$ and $Q_{1}^{\perp}=I-Q_{1}$ denote the respective 
projections onto $\mathbb{C}w_{1}$ and $w_{1}^{\perp}$, then 
\[
  Q_{1}TQ_{1}=aQ_{1},
\]
\[
  Q_{1}^{\perp}TQ_{1}=|\eta^{\perp} \rangle \langle w_{1}|,
\]
\[
  Q_{1}TQ_{1}^{\perp}= 0, \text{ and}
\]
\[
  Q_{1}^{\perp}TQ_{1}^{\perp}= G.
\]

Using now assumptions in theorem (1) and (2), we can conclude that the operator
$a-G$ is invertible with bounded inverse 
\[
  (a-G)^{-1}: (w_{1})^{\perp} \to (w_{1})^{\perp}.
\]

We now turn to powers of operator $T$.  An induction yields the following 
matrix representation:
\[
  T^{n}=
  \begin{pmatrix}
    a^{n} &|& 00 \cdots \\
    ---------- &|& ---\\ 
    (a^{n}-G^{n})(a-G)^{-1}\eta^{\perp} &|& G^{n}
  \end{pmatrix}.
\]
Finally an application of assumption (3) yields the following operator limit
\[
  a^{-n}T^{n} \underset{n \to \infty}{\longrightarrow} 
  \begin{pmatrix}
    1 &|& 00 \cdots \\
    ------&|& ---\\
    (a-G)^{-1} \eta^{\perp} &|& 00 \cdots 
  \end{pmatrix}.
\]

We used that $\eta^{\perp} \in \overline{R(a-T)}$, and that 
\[
  a^{-n}(a^{n}-G^{n})(a-G)^{-1}\eta^{\perp} 
  = (1-a^{-n}G^{n})(a-G)^{-1}\eta^{\perp}
  \underset{n \to \infty}{\longrightarrow}(a-G)^{-1}\eta^{\perp}.
\]
Further, if we set $\xi:=w_{1}+(a-G)^{-1}\eta^{\perp}$, then 
\[
  T\xi=aw_{1}+ \eta^{\perp}+G(a-G)^{-1}\eta^{\perp}
  = aw_{1} + a(a-G)^{-1}\eta^{\perp} = a\xi.
\]
Finally, note that 
\[
  (a-G)^{-1}\eta^{\perp} \in \overline{R(a-T)}=(w_{1})^{\perp}.
\]

It is now immediate from this that all of the statements in the conclusion of
the theorem including (\ref{E:limaT}), (\ref{E:w}) and (\ref{E:xiw}) are satisfied for
the two vectors $w_{1}$ and $\xi$.
\end{proof}

\section{Weighted Frames and Weighted Frame-Operators} 
\label{sec:7}
In this section we address that when frames are considered in infinite-dimensional
separable Hilbert space, then the trace-class condition may not hold.

There are several remedies to this, one is the introduction of a certain weighting
into the analysis.  Our weighting is done as follows in the simplest case: Let
$(h_{n})_{n \in \mathbb{N}}$ be a sequence of vectors in some fixed Hilbert 
space, and suppose the frame condition from Definition \ref{D:frame} is 
satisfied for all $f \in \mathcal{H}$
We say that $(h_{n})$ is a frame.  As in section \ref{sec:5}, we introduce the 
analysis operator $L$:
\[
  \mathcal{H} \ni f \longmapsto (\langle h_{n}|f \rangle)_{n} \in \mathit{l}^{2}
\]
and the two operators
\begin{equation}
\label{E:GopL}
  G:=L^{*}L: \mathcal{H} \to \mathcal{H}
\end{equation}
and
\begin{equation}
\label{E:Gramian}
  G_{R}:=LL^{*}: \mathit{l}^{2} \to \mathit{l}^{2},
\end{equation}
(the Grammian).

As noted,
\begin{equation}
\label{E:Gh}
  G= \sum_{n \in \mathbb{N}}|h_{n} \rangle \langle h_{n}|
\end{equation}
and $G_{R}$ is matrix-multiplication in $\mathit{l}^{2}$ by the matrix 
$(\langle h_{i} | h_{j} \rangle)$, i.e., 
\[
  \mathit{l}^{2} \ni x=(x_{i}) \mapsto (G_{R}x)=y=(y_{i})
\]
where
\begin{equation}
\label{E:yj}
  y_{j} = \sum_{i} \langle h_{j}| h_{i} \rangle x_{i}
\end{equation}

\begin{proposition}
Let $\{h_{n}\}$ be a set of vectors in a Hilbert space
(infinite-dimensional, separable), and suppose these vectors form a
frame with frame-bounds $c_{1}$, $c_{2}$.
\begin{itemize}
\item[(a)] Let $(v_{n})$ be a fixed sequence of scalars in $\mathit{l}^{2}$. 
Then the frame operator $G = G_{v}$ formed from the weighted sequence 
$\{v_{n} h_{n}\}$ is trace-class.
\item[(b)] If $\sum_{n=1}|v_{n}|^{2} = 1$, then the upper frame bound for 
$\{v_{n} h_{n}\}$ is also  $c_{2}$.
\item[(c)] Pick a finite subset $F$ of the index set, typically the
natural numbers $\mathbb{N}$, and then pick $(v_{n})$ in $\mathit{l}^{2}$ such 
that  $v_{n} = 1$ for all $n$ in $F$.
Then on this set $F$ the weighted frame agrees with the initial system
of frame vectors $\{h_{n}\}$, and the weighted frame operator  $G_{v}$ is not
changed on $F$.
\end{itemize}
\end{proposition}
\begin{proof}
(a) Starting with the initial frame $\{h_{n}\}_{n \in \mathbb{N}}$ we form the
weighted system $\{v_{n}h_{n}\}$.  The weighted frame operator arises from 
applying (\ref{E:Gh}) to this modified system, i.e., 
\begin{equation}
\label{E:Gv}
  G_{v} = \sum_{n \in \mathbb{N}}|v_{n}h_{n} \rangle \langle v_{n}h_{n}|
        = \sum_{n \in \mathbb{N}}|v_{n}|^{2}|h_{n} \rangle \langle h_{n}|
\end{equation}
Let $(\epsilon_{n})$ be the canonical ONB in $\mathit{l}^{2}$, i.e., 
$(\epsilon_{n})_{k} := \delta_{n,k}$.  Then $h_{n}=L^{*}\epsilon_{n}$, so
\[
  \|h_{n}\| \leq \|L^{*}\|\|\epsilon_{n}\|=\|L^{*}\|=\|L\|.
\]
Now apply the trace to (\ref{E:Gv}): Suppose 
$\|(v_{\epsilon})\|_{\mathit{l}^{2}}=1$.  Then 
\[
  trG_{v}=\sum_{n \in \mathbb{N}}|v_{n}|^{2}tr|h_{n} \rangle \langle h_{n}|
         = \sum_{n \in \mathbb{N}}|v_{n}|^{2}\|h_{n}\|^{2}
  \leq \|L\|^{2}\|(v_{n})\|_{\mathit{l}^{2}}^{2}
\]
\[
=\|L\|^{2}=\|LL^{*}\|
  =\|L^{*}L\|=\|G\| = sup(spec(G)).
\]
(Note that the estimate shows more: The sum of the eigenvalues of $G_{v}$ is 
dominated by the top eigenvalue of $G$.) 
But we recall (\ref{S:Suppl}) that $(h_{n})$ is a frame with frame-bounds $c_{1}$,
$c_{2}$.  It follows (\ref{S:Suppl}) that $spec(G) \subset [c_{1}, c_{2}]$.  This
holds also if $c_{1}$ is the largest lower bound in (\ref{E:framebd}), and $c_{2}$
the smallest upper bound; i.e., the optimal frame bounds.  

Hence $c_{2}$ is the spectral radius of $G$, and also $c_{2}=\|G\|$.  The 
conclusion in (a)-(b) follows.

(c) The conclusion in (c) is a immediate consequence, but now
\[
  trG_{v} = \sum_{n \in \mathbb{N}}|v_{n}|^{2}\|h_{n}\|^{2}
  \leq \|G\|\sum_{n \in \mathbb{N}}|v_{n}|^{2} 
  = c_{2}(\# F+\sum_{n \in \mathbb{N}\setminus F}|v_{n}|^{2})
\]
where $\# F$ is the cardinality of the set specified in (c).
\end{proof}

\begin{remark}
Let $\{h_{n}\}$ and $(v_{n}) \in \mathit{l}^{2}$ be as in the proposition and let 
$D_{v}$ be the diagonal operator with the sequence $(v_{n})$ down the diagonal.  
Then $G_{v} = L^{*}|D_{v}|^{2}L$, and $G_{R_{v}} = D_{v}^{*}G_{R}D_{v}$; where
\[
|D_{v}|^{2} =D_{v}\overline{D_{v}}=
\begin{pmatrix}
  |v_{1}|^{2} & 0 & 0 & 0 &...&0 &0 &0\\ 
  0 & |v_{2}|^{2} & 0 & 0 &...&0 &0 &0\\
  0 & 0 & \cdot & 0 & 0 &...&0 &0\\
  0 & 0 & 0 & \cdot & 0 &...&0 &0\\ 
  0 & 0 & 0 & 0 & \cdot & 0 &...&0 \\
\end{pmatrix}
\]
\end{remark}

\subsection{$\mathcal{B}(\mathcal{H})=\mathcal{T}(\mathcal{H})^{*}$}
The formula $\mathcal{B}(\mathcal{H})=\mathcal{T}(\mathcal{H})^{*}$ summarizes
the known fact \cite{KaRi97} that $\mathcal{B}(\mathcal{H})$ is a Banach dual 
of the Banach space of all trace-class operators.  

The conditions (\ref{E:cPf}) and (\ref{E:framebd}) which introduce frames, 
(both in vector form and fusion form) may be recast with the use of this 
duality.

\begin{proposition}
\label{P:quantum}
An operator $G$ arising from a vector system $(h_{\alpha}) \subset \mathcal{H}$, or from a projection system $(w_{\alpha}, P_{\alpha})$, yields a frame with
frame bounds $c_{1}$ and $c_{2}$ if and only if 
\begin{equation}
\label{E:trr}
  c_{1}tr(\rho) \leq tr(\rho G) \leq c_{2}tr(\rho)
\end{equation}
for all positive trace-class operators $\rho$ on $\mathcal{H}$.
\end{proposition}
\begin{proof}
Since both (\ref{E:cPf}) and (\ref{E:framebd}) may be stated in the form
\[
   c_{1}\|f\|^{2} \leq \langle f|Gf\rangle \leq c_{2}\|f\|^{2}
\]
and 
\[
  tr(|f \rangle \langle f|)=\|f\|^{2},
\] it is clear that (\ref{E:trr}) is 
sufficient.

To see it is necessary, suppose (\ref{E:cPf}) holds, and that $\rho$ is a 
positive trace operator.  By the spectral theorem, there is a an ONB $(f_{i})$,
and  $\xi_{i} \geq 0$ such that 
\[
  \rho=\sum_{i}\xi_{i}|f_{i} \rangle \langle f_{i}|.
\]
We now use the estimates
\[
   c_{1} \leq \langle f_{i}|Gf_{i}\rangle \leq c_{2}
\]
in 
\[
  tr(\rho G)=\sum_{i}\xi_{i}\langle f_{i}|Gf_{i}\rangle.
\]
Since $tr(\rho)=\sum_{i}\xi_{i}$, the conclusion (\ref{E:trr}) follows.
\end{proof}

\begin{remark}
Since quantum mechanical states (see \cite{KaRi97}) take the form of density
matrices, the proposition makes a connection between frame theory and quantum
states.  Recall, a density matrix is an operator 
$\rho \in \mathcal{T}(\mathcal{H})_{+}$ with $tr(\rho)=1$.

\end{remark}

\section{Localization} 
\label{sec:8}
Starting with a frame $(h_{n})_{n \in \mathbb{N}}$, non-zero vectors index set 
$\mathbb{N}$ for simplicity; see (\ref{E:framebd}), we introduce the operators
\begin{equation}
\label{E:Gs}
  G:=\sum_{n \in \mathbb{N}}|h_{n} \rangle \langle h_{n}|
\end{equation}
\begin{equation}
\label{E:Gvs}
  G_{v}:=\sum_{n \in \mathbb{N}}|v_{n}|^{2}|h_{n} \rangle \langle h_{n}| 
  \quad \text{for } v \in \mathit{l}^{2};
\end{equation}
and the components
\begin{equation}
\label{E:Ghs}
  G_{h_{n}}:= |h_{n} \rangle \langle h_{n}|
\end{equation}

We further note that the individual operators $G_{h_{n}}$ in (\ref{E:Ghs}) are
included in the $\mathit{l}^{2}-$index family $G_{v}$ of (\ref{E:Gvs}). To see 
this, take
\begin{equation}
\label{E:vepsi}
  v=\epsilon_{n}=(0,0,...,0,1,0,...) \quad \text{where 1 is in } n^{th} 
  \text{ place}. 
\end{equation}

It is immediate that the spectrum of $G_{h_{n}}$ is the singleton $\|h_{n}\|^{2}$,
and we may take $\|h_{n}\|^{-1}h_{n}$ as a normalized eigenvector.  Hence for the 
components $G_{h_{n}}$, there are global entropy considerations.  Still in 
applications, it is the sequence of local approximations
\begin{equation}
\label{E:seqappr}
  \sum_{i=1}^{m}\langle \psi_{i} | h_{n} \rangle \psi_{i}=Q_{m}^{\psi}h_{n} 
\end{equation}
which is accessible.  It is computed relative to some $ONB(\psi_{i})$.  The
corresponding sequence of entropy numbers is: 
\begin{equation}
\label{E:Smpsi}
  S_{m}^{\psi}(h_{n}):=-\sum_{i=1}^{m}|\langle \psi_{i}|h_{n} \rangle|^{2}
    \log{|\langle \psi_{i}|h_{n} \rangle|^{2}}
\end{equation}

The next result shows that for every $v \in \mathit{l}^{2}$ with 
$\|v\|_{\mathit{l}^{2}}=1$, the combined operator $G_{v}$ always is 
entropy-improving in the following precise sense.  

\begin{proposition}
\label{P:Smpsi}
Consider the operators $G_{v}$ and $G_{h_{n}}$ introduced in (\ref{E:Gvs}) and 
(\ref{E:Ghs}).  Suppose $v \in \mathit{l}^{2}$ satisfies 
$\|v\|_{\mathit{l}^{2}}=1$.  Then for every ONB $(\psi_{i})$ and for every $m$,
\begin{equation}
\label{E:Smpsineq}
  S_{m}^{\psi}(G_{v})\geq \sum_{n \in \mathbb{N}}|v_{n}|^{2}S_{m}^{\psi}
  (G_{h_{n}}).
\end{equation}
\end{proposition}
\begin{proof}
Let $v$, $\psi$, and $m$ be as specified in the proposition.  Introduce the convex
function $\beta(t):=t\log{t}, \quad t\in [0,1]$ with the convention that
$\beta(0)=\beta(1)=0$.  Then
\[
  -S_{m}^{\psi}(G_{v})
  = \sum_{i=1}^{m}\beta(|\langle \psi_{i}|G_{v}\psi_{i} \rangle|^{2})
  \leq \sum_{i=1}^{m}\sum_{n \in \mathbb{N}}|v_{n}|^{2}\beta
   (|\langle \psi_{i}|G_{h_{n}}\psi_{i} \rangle|^{2})
\]
\[
  =\sum_{n \in \mathbb{N}}|v_{n}|^{2}\sum_{i=1}^{m}\beta
   (|\langle \psi_{i}|G_{h_{n}}\psi_{i} \rangle|^{2})
  =-\sum_{n \in \mathbb{N}}|v_{n}|^{2}S_{m}^{\psi}(G_{h_{n}})
\]
where we used that $\beta$ is convex.  In the last step, formula (\ref{E:Smpsi}) 
was used.  This proves (\ref{E:Smpsineq}) in the proposition.
\end{proof}

\section{Engineering Applications} 
\label{sec:9}
In wavelet image compression, wavelet decomposition is performed on a digital 
image.  Here, an image is treated as a matrix of functions where the entries are
pixels.  The following is an example of a representation for a digitized image 
function:
\begin{equation}
\label{E:imagematrix}
  \mathbf{f(x,y)} =
  \begin{pmatrix}
    f(0,0) & f(0,1) & \cdots & f(0, N-1) \\
    f(1,0) & f(1,1) & \cdots & f(1, N-1) \\
    \vdots & \vdots & \vdots & \vdots \\
    f(M-1,0) & f(M-1,1) & \cdots & f(M-1,N-1)
  \end{pmatrix}.
\end{equation}
After the decomposition quantization is performed on the image, the 
quantization may be a lossy (meaning some information is being lost) or lossless.  Then a 
lossless means of compression, entropy encoding, is done on the image to 
minimize
the memory space for storage or transmission.  Here the mechanism of entropy will be discussed.

\subsection{Entropy Encoding}
\label{sec:10}
In most images their neighboring pixels are correlated and thus contain redundant information.  Our task is to find less correlated representation of the image, then perform redundancy reduction and irrelevancy reduction. Redundancy
reduction removes duplication from the signal source (for instance a digital 
image). Irrelevancy reduction omits parts of the signal that will not be 
noticed by the Human Visual System (HVS).

Entropy encoding further compresses the quantized values in a lossless manner 
which gives better compression in overall. It uses a model to accurately 
determine the probabilities for each quantized value and produces an 
appropriate code based on these probabilities so that the resultant output 
code stream will be smaller than the input stream. 

\subsubsection{Some Terminology}
\label{sec:11}
  \begin{itemize}
    \item[(i)]Spatial Redundancy refers to correlation between neighboring 
    pixel values.  
    \item[(ii)]Spectral Redundancy refers to correlation between different 
    color planes or spectral bands. 
  \end{itemize}

\subsection{The Algorithm}
\label{sec:Alg}
Our aim is to reduce the number of bits needed to represent an image by removing redundancies as much as possible.

The algorithm for entropy encoding using Karhunen-Lo\`{e}ve expansion can be described as follows:
  \begin{itemize}
   \item[1.] Perform the wavelet transform for the whole image. (ie. wavelet decomposition.)
   \item[2.] Do quantization to all coefficients in the image matrix, except the average detail.
   \item[3.] Subtract the mean: Subtract the mean from each of the data dimensions. This produces a data set whose mean is zero.
   \item[4.] Compute the covariance matrix. 
   \[
     cov(X,Y)=\frac{\sum_{i=1}^{n}(X_{i}-\bar{X})(Y_{i}-\bar{Y})}{n}
   \]
  \item[5.] Compute the eigenvectors and eigenvalues of the covariance matrix.
  \item[6.] Choose components and form a feature vector(matrix of vectors), 
  \[(eig_{1}, ... ,eig_{n})\]
Eigenvectors are listed in decreasing order of the magnitude of their 
eigenvalues.  Eigenvalues found in step 5 are different in values.  The 
eigenvector with highest eigenvalue is the principle component of the 
 data set.  
  \item[7.] Derive the new data set. 
  \[
     \text{Final Data} = \text{Row Feature Matrix} \times 
     \text{Row Data Adjust}.
  \]
  \end{itemize} 
Row Feature Matrix is the matrix that has the eigenvectors in its rows 
with the most significant eigenvector (i.e., with the greatest eigenvalue) at
the top row of the matrix.  
Row Data Adjust is the matrix with mean-adjusted data transposed. That is, 
the matrix contains the data items in each column with each row having
a separate dimension.\cite{Smi02}


Inside the paper we use $(\phi_{i})$ and $(\psi_{i})$ to denote generic ONBs 
for a Hilbert space.  
However, in wavelet theory, \cite{Dau92} it is traditional to reserve
$\phi$ for the father function and $\psi$ for the mother function.
A 1-level wavelet transform of an $N \times M$ image can be represented as
\begin{equation}
\label{E:ahvd1}
  \mathbf{f} \mapsto
  \begin{pmatrix}
    \mathbf{a}^{1} & | & \mathbf{h}^{1} \\
    -- & & -- \\
    \mathbf{v}^{1} & | & \mathbf{d}^{1}
  \end{pmatrix} \\
\end{equation}
where the subimages $\mathbf{h}^{1}, \mathbf{d}^{1}, \mathbf{a}^{1}$ and
$\mathbf{v}^{1}$ each have the dimension of $N/2$ by $M/2$.

\begin{equation}
\label{E:ahvdrel}
  \begin{array}{l}
  \mathbf{a}^{1} = V_{m}^{1} \otimes V_{n}^{1} : \phi^{A}(x,y) = \phi(x)\phi(y) 
  = \sum_{i}\sum_{j}h_{i}h_{j}\phi(2x-i)\phi(2y-j) \\
  \mathbf{h}^{1} = V_{m}^{1} \otimes W_{n}^{1} : \psi^{H}(x,y) = \psi(x)\phi(y)
  = \sum_{i}\sum_{j}g_{i}h_{j}\phi(2x-i)\phi(2y-j) \\
  \mathbf{v}^{1} = W_{m}^{1} \otimes V_{n}^{1} : \psi^{V}(x,y) = \phi(x)\psi(y)
  = \sum_{i}\sum_{j}h_{i}g_{j}\phi(2x-i)\phi(2y-j) \\
  \mathbf{d}^{1} = W_{m}^{1} \otimes W_{n}^{1} : \psi^{D}(x,y) = \psi(x)\psi(y)
  = \sum_{i}\sum_{j}g_{i}g_{j}\phi(2x-i)\phi(2y-j)
  \end{array}
\end{equation}
where $\phi$ is the father function and $\psi$ is the mother function in 
sense of wavelet, $V$ space denotes the average space and the $W$ spaces are 
the difference space from multiresolution analysis (MRA) \cite{Dau92}.  
Note that, on the very right hand side, in each of the 
four system of equations, we have the affinely transformed function $\phi$ 
occurring both places under each of the the double summations.
Reason: Each of the two functions, father function $\phi$, and mother function 
$\psi$ satisfies a scaling relation. So both $\phi$ and $\psi$ are expressed 
in terms of the scaled family of functions $\phi(2 \cdot -j)$ with $j$ ranging 
over $Z$, and the numbers $h_{i}$ are used in the formula for $\phi$, while 
the numbers $g_{j}$ are used for $\psi$. 
Specifically, the relations are:
\[
  \phi(x)=\sum_{i}h_{i}\phi(2x-i), \text{ and } 
  \psi(x)=\sum_{j}g_{j}\phi(2x-j).
\]
$h_{i}$ and $g_{j}$ are low-pass and high-pass filter coefficients respectively.
$\mathbf{a}^{1}$ denotes the first averaged image, which consists of 
average intensity values of the original image.  Note that only $\phi$ 
function, $V$ space and $h$ coefficients are used here. 
$\mathbf{h}^{1}$ denotes the first detail image of horizontal components, which consists of intensity difference along the vertical axis of the original 
image. Note that $\phi$ function is used on $y$ and $\psi$ function on 
$x$, $W$ space for $x$ values and $V$ space for $y$ values; and both $h$ and 
$g$ coefficients are used accordingly.
$\mathbf{v}^{1}$ denotes the first detail image of vertical components, which 
consists of intensity difference along the horizontal axis of the original 
image. Note that $\phi$ function is used on $x$ and $\psi$ function on 
$y$, $W$ space for $y$ values and $V$ space for $x$ values; and both $h$ and 
$g$ coefficients are used accordingly. 
$\mathbf{d}^{1}$ denotes the first detail image of diagonal components, which consists of intensity difference along the diagonal axis of the original image. The original image is reconstructed from the decomposed image by taking the sum 
of the averaged image and the detail images and scaling by a scaling factor. It
could be noted that only $\psi$ function, $W$ space and $g$ coefficients are 
used here. 
See \cite{Wal99}, \cite{So06}.

This decomposition is not only limited to one step, but it can be done again and 
again on the averaged detail depending on the size of the image.  Once it 
stops at certain level, quantization (see \cite{Smi02}, \cite{SkChEb01}, 
\cite{Use01}) is done on the image.  This quantization step may be lossy or 
lossless.  Then the lossless entropy encoding is done on 
the decomposed and quantized image.  

There are various means of quantization and one commonly used one is called
thresholding. Thresholding is a method of data reduction where it puts $0$ for
the pixel values below the thresholding value or something other `appropriate'
value.  Soft thresholding is defined as follows:  
\begin{equation}
  T_{soft}(x)=
  \begin{cases}
    0  &\text{if $|x| \leq \lambda$} \\
    x - \lambda &\text{if $x > \lambda$} \\
    x + \lambda &\text{if $x < -\lambda$} 
  \end{cases}
\label{eq-6}
\end{equation}
and hard thresholding as follows:
\begin{equation}
  T_{hard}(x)=
  \begin{cases}
    0  &\text{if $|x| \leq \lambda$} \\
    x  &\text{if $|x| > \lambda$} 
  \end{cases}
\label{eq-7}
\end{equation}
where $\lambda \in \mathbb{R}_{+}$ and $x$ is a pixel value.  It could be 
observed by looking at the definitions, the difference between them is related 
to how the coefficients larger than a threshold value $\lambda$ in absolute 
values are handled. In hard thresholding, these coefficient values are left 
alone. Where else, in soft thresholding, the coefficient values area decreased 
by $\lambda$ if positive and increased by $\lambda$ if negative \cite{Waln02}. 
Also, see \cite{Wal99}, \cite{GoWoEd04}, \cite{So06}. 

Starting with a matrix representation for a particular image, we then compute 
the covariance matrix using the steps from (3) and (4) in algorithm above. 
Next, we compute the Karhunen-Lo\`{e}ve eigenvalues.  As usual, we arrange the 
eigenvalues in decreasing order. The corresponding eigenvectors are arranged to
match the eigenvalues with multiplicity. The eigenvalues mention here are the 
same eigenvalues in Theorem \ref{T:smallest_error} and Theorem 
\ref{T:smallest_entropy}, thus yielding smallest error and smallest entropy in 
the computation.

The Karhunen-Lo\`{e}ve transform or Principal Components Analysis (PCA) allows 
us to better represent each pixels on the image matrix with the smallest 
number of bits.  It enables us to assign the smallest number of bits for the 
pixel that has the highest probability, then the next number to the pixel 
value that has second highest probability, and so forth; thus the pixel that 
has smallest probability gets assigned the highest value among all the other 
pixel values. 

An example with letters in the text would better depict how the mechanism 
works.  Suppose we have a text with letters a, e, f, q, r with the following 
probability distribution:
\begin{center}
  \begin{tabular}{|c|c|}
    \hline
    Letter & Probability \\
    \hline
    a & 0.3\\
    \hline
    e & 0.2\\
    \hline
    f & 0.2\\
    \hline
    q & 0.2\\
    \hline
    r & 0.1\\ \hline
  \end{tabular}
\end{center}
Shannon-Fano entropy encoding algorithm is outlined as follows:
\begin{itemize}
  \item List all letters, with their probabilities in decreasing order of their probabilities.
  \item Divide the list into two parts with approximately equal probability 
(i.e., the total of probabilities of each part sums up to approximately 0.5). 
  \item For the letters in the first part start the code with a 0 bit and 
for those in the second part with a 1.
  \item Recursively continue until each subdivision is left with just one 
letter \cite{BeClWi90}.
\end{itemize}

Then applying the Shannon-Fano entropy encoding scheme on the above table 
gives us the following assignment.
\begin{center}
  \begin{tabular}{|c|c|c|}
    \hline
    Letter & Probability & code\\
    \hline
    a & 0.3 & 00\\
    \hline
    e & 0.2 & 01\\
    \hline
    f & 0.2 & 100\\
    \hline
    q & 0.2 & 101\\
    \hline
    r & 0.1 & 110\\ \hline
  \end{tabular}
\end{center}
Note that instead of using 8-bits to represent a letter, 2 or 3-bits are being 
used to represent the letters in this case.

\subsection{Benefits of Entropy Encoding}
One might think that the quantization step suffices for compression.  It is 
true that the quantization does compress the data tremendously.  After the 
quantization step many of the pixel values are either eliminated or replaced 
with other suitable values.  However, those pixel values are still represented 
with either 8 or 16 bits. See \ref{sec:DIC}. So we aim to minimize the number 
of bits used by means of entropy encoding.  Karhunen-Lo\`{e}ve transform or 
PCAs makes it possible to represent each pixel on the digital image with the
least bit representation according to their probability thus yields the lossless
optimized representation using least amount of memory.  

\begin{acknowledgements} We thank the members of WashU Wavelet Seminar, 
Professors Simon Alexander, Bill Arveson, Dorin Dutkay, David Larson, 
Peter Massopust and Gestur Olafsson for helpful discussions,   
Professor Brody Johnson for suggesting \cite{Wat65}, and Professor Victor 
Wickerhauser for suggesting \cite{Ash90}, and our referee for excellent 
suggestions.
\end{acknowledgements}

\end{document}